\documentclass[twocolumn,showpacs,preprintnumbers,amsmath,amssymb]{revtex4}
\usepackage{graphicx}
\usepackage{bm}

\newcommand{\ybco}{YBa$_2$Cu$_3$O$_{7-\delta}$ }
\newcommand{\ybcos}{YBa$_2$Cu$_3$O$_{6.97}$ }
\begin{document}

\title{Pinning and Dynamics of Magnetic Flux Moving Across the Twin Planes in \ybcos Single Crystal}

\author{A. V. Bondarenko}
 \email{Aleksandr.V.Bondarenko@univer.kharkov.ua}
\author{M. A. Obolenskii}
\author{A. A. Prodan}
\author{A. A. Zavgorodniy}
\address{Physical department, V.N. Karazin Kharkov National
University, 4 Svoboda Square, 61077 Kharkov, Ukraine.
}%

\author{L. M. Fisher}
\address{All-Russian Electrical Engineering Institute, 12
Krasnokazarmennaya Street, 111250 Moscow, Russian Federation.
}%

\author{A. G. Sivakov}
\address{B.I. Verkin Institute for Low Temperature Physics and
Engineering, Ukrainian Academy of Science, 47 Lenin Avenue, 61164
Kharkov, Ukraine.
}%

\author{V. A. Yampol'skii}
\address{A.Ya. Usikov Institute for Radiophysics and Electronics,
Ukrainian Academy of Science, 12 Proskura Street, 61085 Kharkov,
Ukraine.
}%
\date{\today}

\begin{abstract}
We present the transport studies of field variation of the pinning
force for the flux motion across the twin boundaries (TB's) in
\ybcos single crystals. It is found that the depinning current
$J_c$ decreases with an increase in the magnetic field due to
reduction of the portion of vortex lines trapped by the TB's.
However, at transport currents $J\ll J_c$, the vortex velocity
weakly decreases with the increased field, i.e. it is determined
mainly by the release of the vortex lines from the TB's.
\end{abstract}

\maketitle


Pinning and dynamics of the flux-line-lattice (FLL) in the
presence of various defect structures is a subject of long-term
interest. In the \ybco superconductor oxygen vacancies constitute
randomly distributed point-like pinning centers, while twin
boundaries (TB's) constitute correlated plane-like pinning centers
that are aligned along the $c$-axis. Though the number of twins is
much smaller than that of oxygen vacancies, the pinning by TB's
can be strong if vortices are aligned along the plane of twins and
the Lorentz force is non-collinear to the TB's.

The magnetooptic \cite{Duran92,Belyaeva93}, transport
\cite{Chabanenko99,Pastoriza00} and simulation
\cite{Groth96,Crabtree96} studies have demonstrated that, in
magnetic field \textbf{H}$\parallel$\textbf{c}$\parallel$TB's, the
flux moves predominantly along the TB's rather than along the
Lorentz force direction. This guided motion arises due to
different pinning mechanisms for the vortices moving along and
across the plane of twins. For the parallel motion it is governed
by the presence of a random potential within the TB's, while for
the perpendicular motion it is determined by suppressing the
superconducting order parameter at the TB's \cite{Blatter94}. This
implies that the pinning must be much stronger for the
perpendicular motion of the vortices compared with that for the
parallel motion.

Single crystals were grown in a gold crucible by a self-flux
method \cite{Obolenskii90}. Two bridges, B1 and B2, were cut out
from the crystals by a pulsed laser technique \cite{Zhuravel96}.
The length of the measured part of the bridges was 0.5 mm and its
width was 0.2 mm. Final oxygenation of both bridges was made in an
oxygen atmosphere at 400 $^\circ$C for one week. The critical
temperature of bridges B1 and B2 was 92.9 K and 92.7 K,
respectively. The TB's inside the measured part were aligned in
one direction. The distance $d$ between the TB's was about 0.5
$\mu$m and 1 $\mu$m in the bridges B1 and B2, respectively. The
transport dc-current was applied along the $ab$-plane, and it was
parallel to the TB's in the bridge B1 and perpendicular to the
TB's in the bridge B2. Measurements were performed at reduced
temperature $t = T/T_c =$ 0.948 in the field
\textbf{H}$\parallel$\textbf{c}. Thus the Lorentz force was
perpendicular to the TB's in the bridge B1, as shown in the inset
of Fig. 1a, and it was parallel to the TB's in the bridge B1, as
shown in the inset of Fig. 3a. The current contacts area of 2
mm$^2$ allowed to pass dc-current up to 1 A without overheating of
the contacts.

Special attention was paid to the heating effects. The heating was
estimated from the small upward deviation in the $E-J$ curves
measured in the normal state, $t$=1.02, and in the superconducting
state ($t$=0.98, $H$=15 kOe) where the $E(J)$ curves are linear in
absence of the heating. The heating effects at the largest
dissipation level 14~$\mu$W has been estimated 
to be less than 10 mK. We also looked for, but did not observe,
hysteresis in the $E(J)$ curves measured with increasing and
decreasing current.

\begin{figure}
\includegraphics{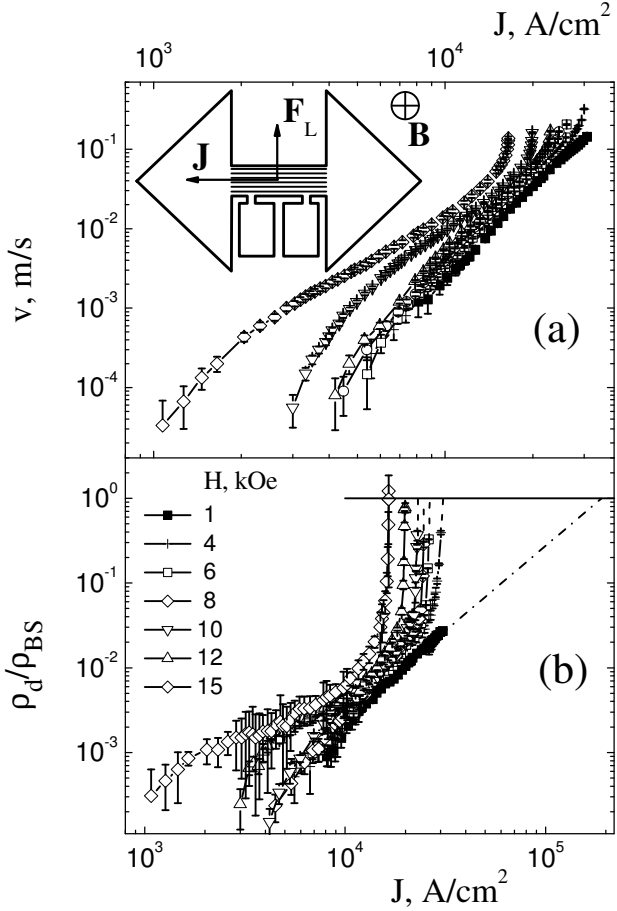}
\caption{\label{fig:1} Current variation of vortex velocity (a)
and differential resistance (b) for the vortex motion across the
TB's. The inset shows sketch of the bridge B1 and geometry of
measurements.}
\end{figure}

\begin{figure}
\includegraphics{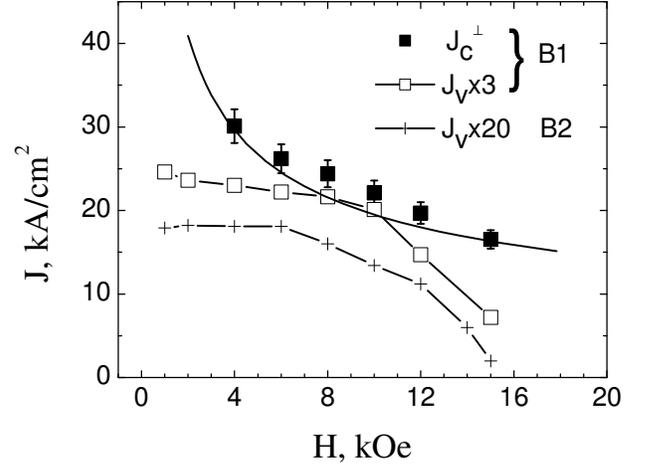}
\caption{\label{fig:2} Field variation of the depinning current
$J_c$ and of the threshold current $J_v$ determined at vortex
velocity $v = 10^{-3}$ m/s.}
\end{figure}

Results of measurements of the bridge B1 are presented in
Fig.~\ref{fig:1}. Panel (a) show the current variation in the
vortex velocity $v(J) = E(J)/cB$, derived from the measured
$E(J)$-curves, and panel (b) shows the current variation of
differential resistance $\rho_d \equiv dE/dJ$, normalized to the
flux-flow resistance $\rho_{ff} = \rho_NB/B_{c2}$
\cite{Bardeen65}. At low currents, the ratio
$\rho_d/\rho_{ff}\ll1$ corresponds to realization of the flux
creep regime, and at high currents the ratio $\rho_d/\rho_{ff}$
quickly approaches 1 indicating onset of the flux depinning.
Extrapolating the ratio $\rho_d/\rho_{ff}$ (which corresponds
onset of the depinning) to one, as shown in the Fig.~\ref{fig:1}b
by the dashed lines, we obtained the field variation of the
depinning current $J_c^\perp(H)$ shown in Fig.~\ref{fig:2}. It is
seen that the depinning current decreases with the increased field
and, thus, the vortex velocity and ratio $\rho_d/\rho_{ff}$
increases rapidly with the field when the current approaches the
value of $J_c^\perp(H)$.

In the studied geometry, the pinning potential is heavily
non-uniform along the Lorentz force action. The weak random point
potential constituted by oxygen vacancies is modulated by the
strong 2D potential of the TB's. In the investigated field region,
1 kOe $\leq H \leq$ 15 kOe, the intervortex distance $a_0 =
(\Phi_0/B)^{1/2} = 40\div150$ nm is smaller compared with the
intertwins separation. Therefore, the fraction of vortices trapped
by the TB's, $n_{TB} \cong a_0/d$, is smaller than that of the
vortices placed in between the TB's, $n_b \cong (1 - a_0/d)$.
Accordingly, most of vortices are weakly pinned by point defects
and the insignificant part of them is strongly pinned by the TB's.

\begin{figure}
\includegraphics{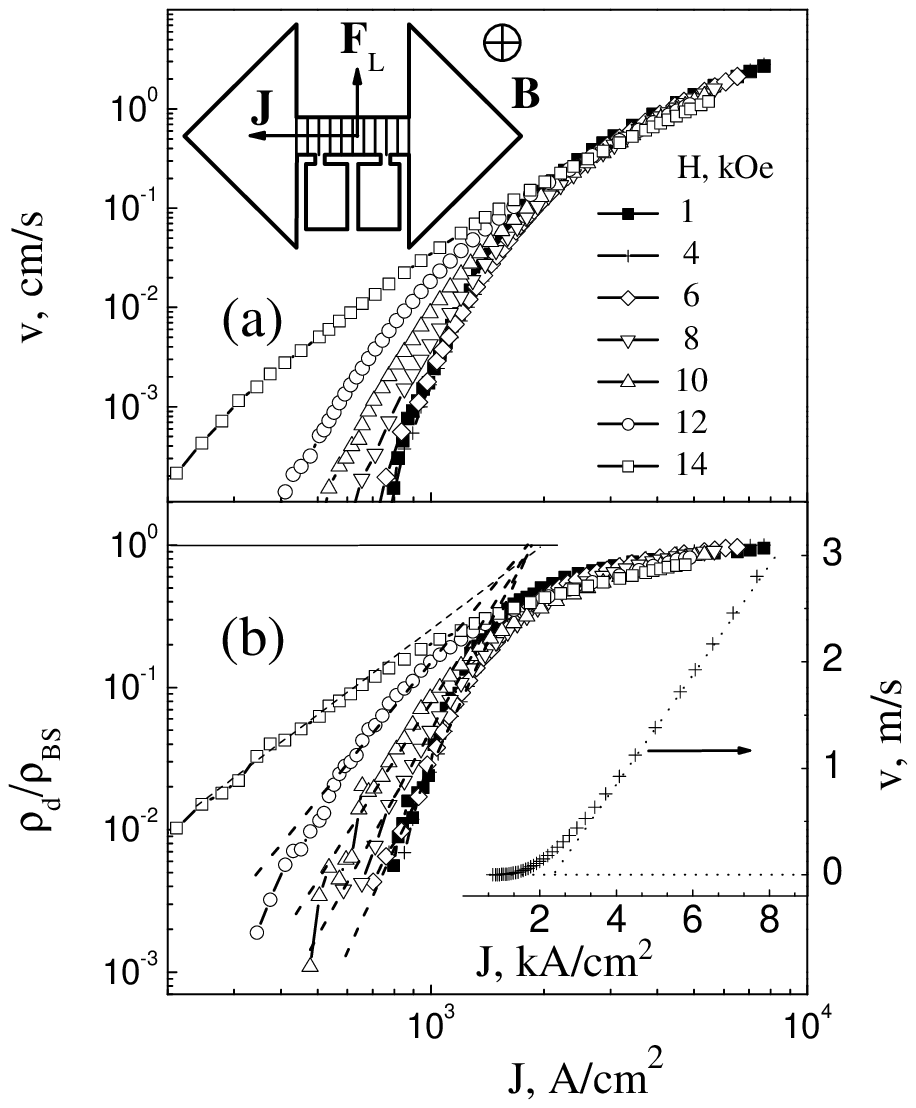}
\caption{\label{fig:3} Current variation of vortex velocity (a)
and differential resistance (b) for the vortex motion along the
TB's. The inset of panel (a) shows sketch of the bridge B2 and
geometry of measurements. The inset of panel (b) shows the
$v(J)$-curve for $H$ = 4 kOe and a method of estimation of the
depinning current $J_c^b$.}
\end{figure}

In order to estimate the contribution of TB's into the total
pinning force, it is necessary to know the pinning force for the
vortices within the bulk of the crystal. Such information was
obtained from measurements of the bridge B2. In this bridge the
Lorentz force is directed along the TB's, and the pinning arises
due to interaction of vortices with the random point potential.
The obtained $v(J)$ and $\rho_d(J)/\rho_{ff}$ dependences are
shown in Fig.~\ref{fig:3}a and in Fig.~\ref{fig:3}b, respectively.
At low currents, the ratio $\rho_d/\rho_{ff}$ is very small, but
at high currents the value of  $\rho_d/\rho_{ff}$ is around unity,
indicating that the measurements were performed in the flux creep
and flux flow regimes. As evident from Fig.~\ref{fig:3}a in
magnetic fields $H\leq6$kOe, the velocity $v$ does not depend on
the field within both regimes of vortex motion. The depinning
current can be estimated by a  conventional method, i.e.
extrapolating the linear parts of $v(J)$ curves (which corresponds
to the flow regime) to the zero velocity, as is shown in the inset
to Fig.~\ref{fig:3}b, and by extrapolating the ratio
$\rho_d/\rho_{ff}$ (which corresponds the creep regime) to 1 (see
Fig.~\ref{fig:3}b). These methods give the value of the $J_c^b
\cong 2.36 \pm 0.2$ and $1.95 \pm 0.2$ kA/cm$^2$, respectively.
Taking into account that final oxygenation of the bridges B1 and
B2 was made under the same condition, the point pinning potentials
are identical in both bridges. Thus, we can assume that the
critical current due to the pinning of vortices within the bulk of
the bridge B1 is about 2.2 kA/cm$^2$.

Assuming that the total depinning current is an additive parameter
we can write
\begin{eqnarray}
J_c^\perp\cong(1-n_{TB})J_c^b+n_{TB}J_{TB}^\perp \label{eq:jcperp}
\end{eqnarray}
where $J_{TB}^\perp$ is the depinning current of vortices trapped
by the TB's. The current $J_{TB}^\perp$ is determined by
suppression of the superconducting order parameter at the TB's and
can be written as \cite{Blatter94}
\begin{eqnarray}
J_{TB}^\perp\cong(\varepsilon_{TB}/\varepsilon_0)J_0
\label{eq:jtbperp}
\end{eqnarray}
where $J_0=(4/3\surd\overline{3})(c\varepsilon_0/\xi\Phi_0)$ is
the depairing current, $\varepsilon_0=(\Phi_0/4\pi\lambda)^2$, and
$\varepsilon_{TB}$ is the pinning potential of the TB's. Assuming
$t$ = 0.95, $\lambda(t)$ = 400 nm, and $\xi(t)$ = 6 nm, the
depairing current is estimated to be about $10^7$ A/cm$^2$. The
decoration \cite{Vinnikov90}, magnetooptic \cite{Dorosinskii95},
and transport \cite{Kwok96} experiments give the value of ratio
$\varepsilon_{TB}/\varepsilon_0=0.017\div0.026$, and for
reasonable ratio $\varepsilon_{TB}/\varepsilon_0=0.021$ we obtain
the value of the current $J_{TB}=210$ kA/cm$^2$. Field variation
of the current $J_c$ determined by Eq.~\ref{eq:jcperp} for the
values of $J_c^b =$ 2.2 kA/cm$^2$ and $J_{TB}^\perp =$ 210
kA/cm$^2$ is shown by the solid curve in Fig.~\ref{fig:2}, and it
well describes experimental data. This indicates that depinning
force is really additive parameter, and decrease of the
$J_c^\perp$ is caused by the reduction of fraction of vortex lines
trapped by the TB's.

As is seen from Fig.~\ref{fig:1}, in magnetic fields
$H\leq$10~kOe, the thermally assisted creep across the TB's occurs
at currents $J>$~4 kA/cm$^2$ which exceed the depinning current of
the vortices placed within the bulk of the crystal. This means
that the creep is controlled by the strong pinning of vortices
trapped by the TB's. It should be pointed out that the vortices
trapped by the TB's are subjected not only to the action of the
Lorentz force, but also to the pressure caused by the vortices
placed in between the TB's because the measurement is performed at
currents $J>J_c^b$. Considering that the ratio between the number
of vortices placed in between the TB's and that trapped by the
TB's, $(1-n_{TB})/n_{TB}$, increases with the magnetic field, the
pressure per unit trapped vortex line increases with the magnetic
field too. However, as seen from Fig.~\ref{fig:2}, threshold
current $J_v^\perp$ , determined within the creep regime at vortex
velocity criteria $v = 10^{-3}$ m/s, much weaker decreases with
increased field compared to the $J_c^\perp$. This indicates that
creep is mainly controlled by the release of vortex lines from the
TB's potential wall, but it weak decreases with increased pressure
exerted by the vortices placed in between the TB's. A possible
reason of this is the high longitudinal correlation length $L_c^b=
\varepsilon\xi(J_0/J_c)^{1/2}$ of the vortices within the bulk of
the crystal compared with the longitudinal correlation length
$L_c^{TB} =
\xi(6\varepsilon_l\varepsilon_{TB}/\varepsilon_0^2)^{1/2}(J_0/J)^\mu$
of vortices trapped by the TB's \cite{Blatter94}. Here
$\varepsilon$ is the anisotropy parameter, and $\varepsilon_l =
\varepsilon_0\varepsilon^2$ is the vortex line tension in the
field \textbf{H}$\parallel$\textbf{c}. Indeed, the length $L_c^b$
is about $65\varepsilon \xi$, while the length $L_c^{TB}$ for
experimentally determined exponent $\mu \approx 0.4$ and for
transport currents $J > 4 kA/cm^2$ is less than $7\varepsilon
\xi$. Thus, the vortices trapped by the TB's experience the
external action on the length $L_c^b$, which is about ten times
larger than $L_c^{TB}$ and, consequently, the effect of pressure
is substantially reduced.

It is important to notice difference in the flux dynamics in
bridges B1 and B2. As indicated above, in bridge B2 the current
variations of the ratio $\rho_d/\rho_{ff}$, corresponding to the
creep regime, extrapolate to the value of ratio $\rho_d/\rho_{ff}$
= 1 at approximately the same value of current and this value
coincides within experimental error with the critical current
determined within the flow regime. This indicates that for the
parallel vortex motion the critical current is determined by the
same kind of pinning centers, namely by the point defects. In
contrast, in bridge B1 the current variations of the ratio
$\rho_d/\rho_{ff}$ corresponding to the creep regime extrapolate
to the value of ratio $\rho_d/\rho_{ff}$ = 1 at values of
currents, which substantially exceed the depinning currents.This
supports our conclusion that for perpendicular motion the
depinning and creep of vortices are controlled by the different
kind of the pinning centers. Besides, in the lowest of the studied
field of 1 kOe, when the part of vortices trapped by the TB's is
maximal, the ratio $\rho_d/\rho_{ff}$, corresponding to the creep
regime, extrapolates to the current $190\pm 20 kA/cm^2$ (see the
dash-dot line in Fig.~\ref{fig:1}b), which is close the above
value of $J_{TB}^\perp = 210 kA/cm^2$. This supports assumption
that the current $J_c^\perp$ is really determined by
Eq.~\ref{eq:jcperp}

In conclusion, the results of transport studies of the magnetic
flux dynamics for its motion across the TB's are presented. We
show that the depinning current is the additive characteristic,
and is determined by both pinning at the TB's and pinning by the
point defects. In crystals with a small oxygen deficiency
($\delta\leq$~0.03) the transverse pinning by TB's is almost two
orders of magnitude larger compared with pinning by point defects.
The depinning current decreases with an increased magnetic field
due to the reduced portion of the vortex lines trapped by the
TB's. In contrast, at temperatures not very close to the melting
point of the flux-line-lattice the speed of vortex creep weak
increases with the magnetic field, i.e. it is primarily controlled
by the release of the vortex lines from the TB's potential wall.

The work was supported by INTAS Project 01-2282.


\bibliography{paper}

\end{document}